\documentclass[12pt,a4paper,dvips]{article}

\usepackage{amssymb,amsmath,graphicx,a4p}
\usepackage{cite,mcite}

\usepackage{ifthen}
\usepackage{l3_title}
\usepackage{Lep}

\usepackage{higgs_input}
\journalname{Phys. Lett. B}
\date{October 13, 1999}
\preprint{99-145}
\Lep{2}
\graphicspath{{./figs/}}
\usepackage[nolists,noheads,nomarkers,tablesfirst]{endfloat}
\begin{document}
\begin{titlepage}
  
  \title{Search for Neutral Higgs Bosons of the Minimal Supersymmetric
    Standard Model in $\boldsymbol{\epem}$ Interactions at
    $\boldsymbol{\rts}$ = 189$\boldsymbol{\GeV}$ }

  \author{L3 Collaboration}
  
%
% The abstract
%
\begin{abstract}
  A search for the lightest neutral scalar and neutral pseudoscalar
  Higgs bosons in the Minimal Supersymmetric Standard Model is
  performed using 176.4\pb of integrated luminosity collected by L3 at
  a center-of-mass energy of 189\GeV.  No signal is observed, and the
  data are
  consistent with the expected Standard Model background.  Lower
  limits on the masses of the lightest neutral scalar and pseudoscalar
  Higgs bosons
  are given as a function of $\tanb$.  Lower mass limits for $\tanb
  > 1$ are set at the 95\% confidence level to be $\mh >
  77.1\GeV$ and $\mA > 77.1\GeV$.
%  At low $\tanb$, values of
%$\mh$ less than 96~\GeV{} are excluded, ruling out the region of
% $ 0.8 < \tanb < 1.5 $ in the MSSM.
\end{abstract}
  
%
% Adds "To be submitted to ..." or "Submitted to ...", if relevant
%
\submitted
%\vfill
%\texttt{/hp/pri2b/data/tully/MSSM\_1999/MSSM\_1999.ps}
\end{titlepage}
%
%%%%%%%%%%%%%%%%%%%%%%%%%%%%%%%%%%%%%%%%%%%%%%%%%%%%%%%%%%%%%%%%%%%%%%%%%%%%%%%
% Introduction
%%%%%%%%%%%%%%%%%%%%%%%%%%%%%%%%%%%%%%%%%%%%%%%%%%%%%%%%%%%%%%%%%%%%%%%%%%%%%%%
%
\section*{Introduction}
\label{sec:intro}
In the Minimal Supersymmetric Standard
Model (MSSM)~\cite{mssm_1} two Higgs
doublets  are required, giving rise to
five Higgs bosons: a charged scalar pair, \Hpm; two neutral scalars, \h
and \bigH; and a neutral pseudoscalar, \A.
Within this framework, the \h and \A are predicted to be the lightest Higgs
particles and, therefore, the most likely to be observed at LEP.
The two main production mechanisms are investigated in this letter:

\begin{align}
  &\epemtohZ \label{eq:higgstrahlung} \\
  &\epemtohA. \label{eq:pairproduction}
\end{align}

Process~(\ref{eq:higgstrahlung}) is very similar to the dominant
Standard Model Higgs production mechanism, for which L3 has set a
lower limit on the mass of the Higgs at
95.3\GeV~\cite{l3_sm_higgs_99_paper}.  The production rate
for~(\ref{eq:higgstrahlung}) is, in general, smaller than that of the
Standard Model reaction, but this is compensated by the additional
pair-production process~(\ref{eq:pairproduction}).

Previous searches for the h and A bosons have been reported by L3~\cite{l3_1998_16} and
other
experiments~\cite{opal_4}.
In this letter, our sensitivity to these particles is extended by
including the data taken at $\rts=189\GeV$ and by scanning over a
larger MSSM parameter space.
% At this center-of-mass energy, we have
%good sensitivity over the entire low-\tanb region of the MSSM for all
%standard LEP benchmark mixing scenarios, however, the region
%theoretically allowed in the MSSM is sensitive to the mass of the top
%and to higher order corrections to the theory.
% leaving the usual wiggle room for theorists.

%
%%%%%%%%%%%%%%%%%%%%%%%%%%%%%%%%%%%%%%%%%%%%%%%%%%%%%%%%%%%%%%%%%%%%%%%%%%%%%%%
% Data/MC samples
%%%%%%%%%%%%%%%%%%%%%%%%%%%%%%%%%%%%%%%%%%%%%%%%%%%%%%%%%%%%%%%%%%%%%%%%%%%%%%%
%
\section*{Data and Monte Carlo Samples}
The data were collected using the L3
detector~\cite{l3_1990_1}
at LEP during 1998.  The integrated luminosity is 176.4\pb at an
average center-of-mass energy of 188.7\GeV.

The signal cross sections and branching ratios are calculated using
the HZHA generator~\cite{hzha}.  For the efficiency studies, Monte
Carlo samples of Higgs events are generated using
PYTHIA~\cite{pythia_1} and HZHA.  For the
background studies, the following Monte Carlo programs are used:
PYTHIA (\epemtoqqg), KORALW~\cite{koralw_1} (\epemtoWW),
KORALZ~\cite{koralz} (\epemtotautau),
PHOJET~\cite{phojet_1} (\epemtoeeqq),
EXCALIBUR~\cite{excalibur} (\epemtoffff) and PYTHIA (\epemtoZZ and
$\epem\!\rightarrow\!\Z\epem$).  The number of simulated background
events for the most important background channels is typically 100
times the number of collected data events.  The Monte Carlo signals
are 300 times the number of events expected to be observed with these
luminosities.

The L3 detector response is simulated using the GEANT~3.15
program~\cite{geant}, which takes into account the effects of energy
loss, multiple scattering and showering in the detector. The GHEISHA
program~\cite{gheisha} is used to simulate hadronic interactions in
the detector.

%
%%%%%%%%%%%%%%%%%%%%%%%%%%%%%%%%%%%%%%%%%%%%%%%%%%%%%%%%%%%%%%%%%%%%%%%%%%%%%%%
% Analysis Procedure
%%%%%%%%%%%%%%%%%%%%%%%%%%%%%%%%%%%%%%%%%%%%%%%%%%%%%%%%%%%%%%%%%%%%%%%%%%%%%%%
%
\section*{Analysis Procedures}
The search for \hA and \hZ production is carried out within a constrained
MSSM assuming unification of the scalar fermion masses, the gaugino masses
and the trilinear Higgs-fermion couplings at the GUT scale.
This choice has little impact on the Higgs mass phenomenology but reduces
significantly the number of free parameters. 
%using two
%different sets of values of MSSM parameters, as suggested in
%Reference~\cite{lep2_higgs}.  
The universal scalar fermion mass $m_0$ and the gaugino mass parameter
$M_2$ are fixed to \mbox{1 TeV}. The Higgs mass parameter $\mu$ is set
to $-0.1\TeV{}$. Two extreme scenarios are considered
corresponding to maximal and minimal scalar top mixing as suggested in
\mbox{Reference  ~\cite{lep2_higgs}}. The minimal mixing scenario corresponds
to setting the trilinear Higgs-fermion coupling $A$ to zero. Maximal scalar
top mixing occurs at $A=\sqrt{6}\;\rm{TeV}$. A scan is then performed, in
each mixing scheme, over
the two remaining free parameters \mA and \tanb.  For this search,
the minimum value of
\tanb considered has been decreased from 1.0 to 0.7 and the minimum \A
mass considered has been decreased from 30\GeV to 10\GeV with respect to
our previous publication.  Values of \mA in the range $\mA<10$\GeV
have been previously excluded at LEP~\cite{opal_2}.

The two Higgs production mechanisms, \epemtohA and \epemtohZ, vary in
relative importance as a function of \tanb.  The production of \hA is
dominant at high \tanb, while \hZ production is dominant at low \tanb.
The description of the \hZ analyses at $\rts=189\GeV$ of the decay
modes other than \hZtobbqq and \hZtobbtt can be found in
Reference~\cite{l3_sm_higgs_99_paper}.  The analyses for \hZtobbqq and
\hZtobbtt(\tautau\qqbar) used in this letter have been optimized to
account for the analogous signatures in the \hA channel: \hAtobbbb and
\hAtobbtt.

For values of \mA less than 30\GeV, decays of the \h into a pair of \A bosons
become possible. The \A decays predominantly to b quarks and tau
leptons for most of the \tanb region probed.  The \hZtobbqq analysis
has a significant cross-efficiency for the \hZtoAAff channel
and is used to search for this process.

Common search procedures are applied to both the \hA and \hZ
channels.  First, a preselection is applied which significantly
reduces background while keeping high signal efficiency.  This is
especially effective against background from the two-photon interaction,
which has a large
cross section at these LEP energies.  Second, a final set of selection
cuts is chosen to distinguish signal from background.  Once the final
selection has been applied, a discriminating variable as defined
in \mbox{Reference~\cite{l3_sm_higgs_99_paper,l3_1998_16}} is calculated
for each scan point in the $(\tanb,\mA)$ plane.

There is a significant overlap in the selection for \hA and \hZ in
both the channels involving either four jets, or two jets and two
taus.  The confidence level calculation requires that all events be
uniquely assigned to a given channel.  To this end, for events that
pass both the \hA and \hZ selections, an unique assignment is made based on the
reconstructed masses and the relative production rates at each scan
point.
%which depends on the mass hypothesis and relative production rates of
%\hA and \hZ.  The spectrum of this discriminant is recomputed for each
%point in the $(\tanb,\mA)$ scan and it is used in the confidence level
%calculation which tests for the presence of a signal.

%
%%%%%%%%%%%%%%%%%%%%%%%%%%%%%%%%%%%%%%%%%%%%%%%%%%%%%%%%%%%%%%%%%%%%%%%%%%%%%%%
% 4 jet channel
%%%%%%%%%%%%%%%%%%%%%%%%%%%%%%%%%%%%%%%%%%%%%%%%%%%%%%%%%%%%%%%%%%%%%%%%%%%%%%%
%
\subsection*{The $\boldsymbol{\hAtobbbb}$\, and $\boldsymbol{\hZtobbqq}$\, Channels}
\label{sec:bbbb}
The signature of both the \hAtobbbb and \mbox{\hZtobbqq} decay modes
is four high-multiplicity hadronic jets and the presence of b hadrons.
The dominant backgrounds come from \qqbar production and hadronic
decays of \W pairs and \Z pairs.  In the case of \hAtobbbb, the
identification of b hadrons plays an especially important role.
The analysis follows closely that of Reference~\cite{l3_1998_16}.

%First, a high multiplicity hadronic preselection, common to both \hA
%and \hZ, is applied.  Events are required to have greater than 15
%tracks, a visible energy greater than 60\% of the beam energy, and a
%parallel and perpendicular imbalances less than 30\% of the beam
%energy.  Cuts are applied on the maximum photon and lepton energy to
%reject semi-leptonic W decays and initial state photon events.  The
%combined pre-selection reduces our efficiency on 2 photon background
%to approximately zero.  Adding a loose cut on the $Y_{\mathrm{cut}}$
%parameter in the DURHAM scheme, \Ytf, ensures the four-jet nature of
%the selected events and significantly reduces the background from
%\qqbar events.  Events passing the preselection are then forced to have
%four jets using the DURHAM~\cite{DURHAM} clustering algorithm and a
%kinematic fit requiring 4-momentum conservation (4C) is performed.
First, a high multiplicity hadronic preselection, common to both \hA
and \hZ, is applied which eliminates background from the two-photon
interaction.  The
preselection is similar to the one used at $\rts=183\GeV$
%~\cite{l3_1998_16} 
and only minor changes are made to account  for
the increased center-of-mass energy.  Events passing the preselection
are then forced to have four jets using the DURHAM~\cite{DURHAM}
clustering algorithm, and a kinematic fit requiring four-momentum
conservation (4C) is performed.

Once the preselection has been satisfied, an optimization procedure
%~\cite{l3_1998_16} 
is applied on the Monte Carlo to choose
cuts on variables that maximize the separation between signal and
background.  These optimized cuts serve mainly to reject the multi-jet
QCD background and are dependent on the topology being investigated:
\hA or \hZ.  Selection cuts are placed on the maximum and minimum
dijet mass, minimum jet energy, maximum jet energy difference
and on \Ytf, being the value of
the DURHAM jet resolution
parameter for which the event is transformed from a four-jet to a three-jet
topology.
%(the $Y_{\mathrm{cut}}$
%parameter characterizing four-jet events in the DURHAM scheme).
Values of the cuts for the \hA and \hZ analyses are shown in
Table~\ref{tab:cuts}.  The number of observed and expected events from
Standard Model processes in
the $\rts=189\GeV$ data along with the signal efficiencies for the
preselection and selection cuts are shown in Table~\ref{tab:4jet_eff}.

Events passing the selection cuts are then classified in three
categories: 1) those that pass only the hA cuts; 2) those that pass only 
the hZ cuts; and 3) those that pass both sets of cuts. Category 3) is
then split into two separate samples by choosing the most likely
hypothesis based on the relative production rate for hA and hZ
and the probability of the \mbox{mass $\chi^2$} as defined in
\mbox{Reference ~\cite{l3_1998_16}}.
%Approximately half of the events passing the \hA selection also pass
%the \hZ selection.  This class of events is split into two separate
%samples by choosing the most likely production mechanism at this scan
%point based on the relative production rates and the consistency of
%the reconstructed masses with either $(\mZ,\mh)$, or $(\mA,\mh)$ as
%described in \mbox{Reference ~\cite{l3_1998_16}}.
%All four-jet events now fall into four categories: 1) those passing
%the set of cuts for \hA, 2) those passing \hZ, and those passing both
%but categorized as 3) \hA or 4) \hZ.

\begin{table}[htbp]
  \begin{center}
    \leavevmode
    \begin{tabular}{lll} \hline
      \multicolumn{1}{c}{Cut} & \multicolumn{1}{c}{\hA} & \multicolumn{1}{c}{\hZ} \\
      \hline
      Minimum dijet mass (GeV) & $>$ 15.7 & $>$ 25.3 \\
      Maximum dijet mass (GeV) & $<$ 135.3 & $<$ 118.7 \\
      Minimum jet energy (GeV) & $>$ 15.1 & $>$ 25.9 \\
      Maximum $\Delta E_{\rm jet}$ (GeV) & $<$ 54.8 & $<$ 42.4 \\
      \Ytf & $>$ 0.003 & $>$ 0.009 \\
      Visible Energy (GeV) & $>$ 129.3 & $>$ 133.8 \\
      Number of Tracks & $>$ 25 & $>$ 22 \\
      \hline
    \end{tabular}
    
    \caption{
      Selection cuts for the \hA and \hZ four-jet Higgs search
      channels.  In addition to those abbreviations defined in the text,
the symbol $\Delta E_{\rm jet}$ is the energy difference between any two jets
of the four-jet system.}
    \label{tab:cuts}
  \end{center}
\end{table}

\begin{table}[htb]
  \begin{center}
    \begin{tabular}{crrr}\hline
& \multicolumn{3}{c}{Number of Events} \\
Process & Preselection & \hA cuts & \hZ cuts \\
\hline
\epemtoeeqq & 7.1     & 0.7      & 0.6      \\
\epemtoqq   & 758.0   & 203.7    & 57.4     \\
%\epemtoWW   & 1312.0   & 912.4    & 581.8    \\ ADD TO Wenu
%\epemtoZee  & 16.6     &          &          \\
%$\epem\!\rightarrow \mathrm{q\bar{q}'e\nu}$ 
%            & 19.7     & 1.1      & 0.3      \\
\epemtoWW   & 1331.7   & 913.5    & 582.1    \\
\epemtoZZ   & 76.0     & 47.5     & 37.6     \\
\hline
Total Expected      & 2172.8   & 1165.4   & 677.7    \\
Data        & 2141 \hspace*{5.5pt}   & 1110 \hspace*{5.5pt}   & 641 \hspace*{5.2pt}   \\ \hline \hline
%    \end{tabular}
%    \begin{tabular}{lrrr}
\hspace*{6pt}Efficiency \hAtobbbb $\rule{0pt}{12pt}$
            & 91.5\%   & 77.1\%   & 43.6\%   \\
Efficiency \hZtobbqq   & 93.3\%   & 78.2\%   & 66.2\%   \\
\hline
    \end{tabular}
    \caption{Number of events expected and observed in the 
      four-jets channels.  The signal
      efficiencies at $\rts=189\GeV$ are quoted for hA at $\mA=\mh=80\GeV$
and for hZ at $\mh=95\GeV$.}
    \label{tab:4jet_eff}
  \end{center}
\end{table}

In the final step, the analysis is optimized for four regions in the
\tanbmh plane near the limit of our discovery potential.  For this,
the $\Btag$ variable (Figure 1a), 
the Higgs production angle with respect to the beam axis, $\Theta$, (Figure 1b)
and the probability for
the $\chi^2$ of the Higgs mass hypothesis (Figure 1c) are used. 
The relative discriminating power of these
variables changes with the Higgs mass hypothesis.  For this reason,
a cut optimization is performed at four points in the \tanbmh plane:
(2.7,95\GeV), (7.5,80\GeV), (20,80\GeV) and (50,80\GeV).

%these variables changes for each scan point and different cuts
%values are placed in the four regions.
%production angle variables are optimized for the four analysis
%categories at four regions in the \tanbma plane near the limit of our
%discovery potential.  Distributions of the final discriminant f
%In the final step of the analysis, optimized cuts are placed on the
%b-tag and mass variables shown in Figure~\ref{fig:last_cuts}.  The
%mass variable is the logarithm of the probability of $\chi^2$ assuming
%a signal hypothesis.  The formula for the $\chi^2$ can be found in
%Reference~\cite{l3_1998_16}.  In the case of the \hA categories, a cut
%is also applied to the cosine of the production angle of the Higgs
%bosons with respect to the beam axis.  The spin difference between \W
%and Higgs bosons provides good distinguishing power between the two,
%as can be seen in Figure~\ref{fig:last_cuts}.  

The final discriminating variable is the logarithm of the weighted
combination
%~\cite{l3_1998_16} 
of the probabilities of the $\Btag$ and
the mass $\chi^2$ to be consistent with background.  
%The mass
%variable and production angle vary in discriminating power depending
%on the Higgs mass hypothesis.  For this reason, the relative weight of
%the mass information as well as the cut on the b-tag, mass, and
%production angle variables are optimized for the four analysis
%categories at four regions in the \tanbma plane near the limit of our
%discovery potential.  
Distributions of the final discriminant for the
\hA search and the \hZ search are shown in
Figure~\ref{fig:final_plot}.

\begin{figure}[htb]
  \begin{center}
    \includegraphics*[width=0.9\textwidth,bb=12 201 625 1018]{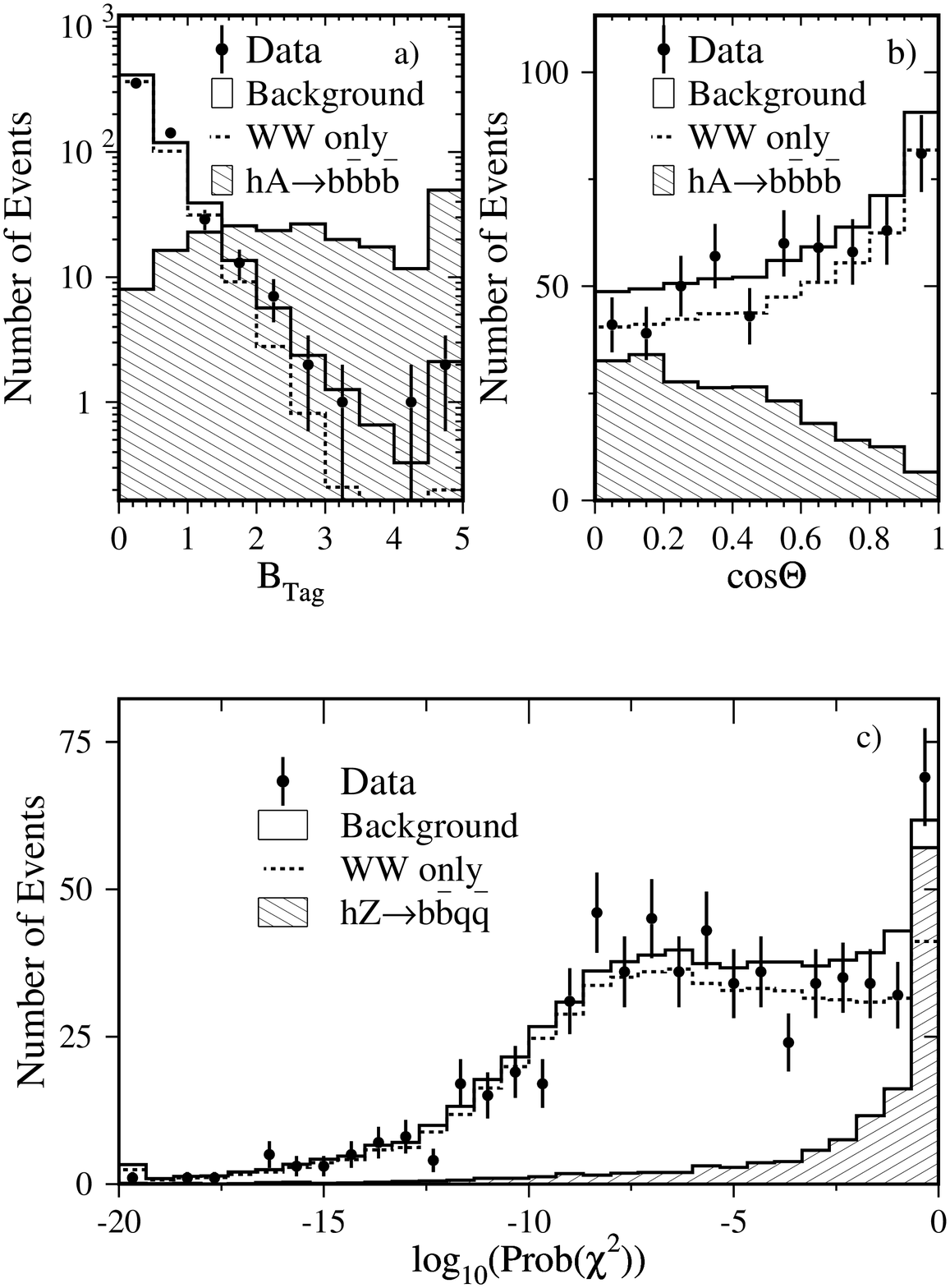}    
    \caption{Distributions of the a) $\Btag$ and b) cosine of the Higgs
      production angle $\Theta$ in the four-jets search.
      The hatched histogram is the expected
      \hA signal (multiplied by a factor of 50) for $\mh=80\GeV$ and $\tanb=50$.
      Distribution c) is the logarithm of the probability of the mass $\chi^2$.
     The hatched histogram is the expected \hZ signal (multiplied by a factor
     of 10) for
     $\mh=95\GeV$ and $\tanb=3$.}
    \label{fig:last_cuts}
  \end{center}
\end{figure}

\begin{figure}[htb]
  \begin{center}
    \includegraphics*[width=0.9\textwidth,bb=8 200 628 1020]{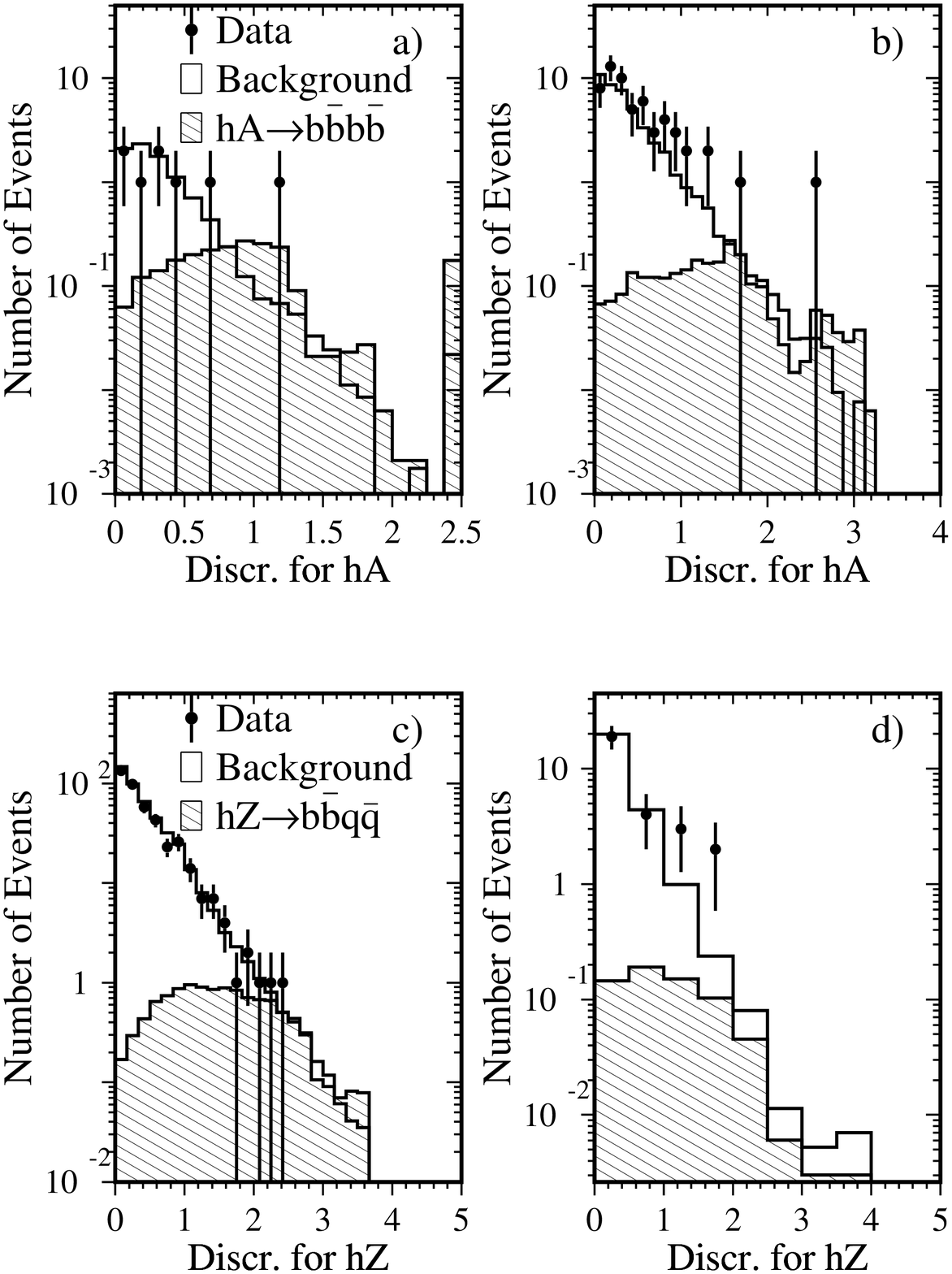}    
    \caption{Distributions of the final discriminant for the
      category of events passing a) both \hA and \hZ cuts but classified as
      \hA candidates and b) events passing only the set of cuts for \hA.
      The hatched histogram is the \hA signal expectation
      for $\mh=80\GeV$ and $\tanb=50$.
      Distributions are plotted for c) the events passing both \hA and
      \hZ cuts but classified as \hZ
      and d) events passing only the \hZ selection.
      The hatched histogram is the \hZ signal expectation for
      $\mh=95\GeV$ and $\tanb=3$.}
    \label{fig:final_plot}
  \end{center}
\end{figure}
%
%%%%%%%%%%%%%%%%%%%%%%%%%%%%%%%%%%%%%%%%%%%%%%%%%%%%%%%%%%%%%%%%%%%%%%%%
% Complicating life with h->AA
%%%%%%%%%%%%%%%%%%%%%%%%%%%%%%%%%%%%%%%%%%%%%%%%%%%%%%%%%%%%%%%%%%%%%%%%
%
\subsection*{The $\boldsymbol{\hZtoAAff}$ Channel}
To investigate \h decays into \A-pairs in the region of very low
\tanb and low \mA, where this channel becomes dominant, the \hZ four
jet analysis described above is employed.  The signature of this
process is at least four hadronic jets with very
high probabilities to contain b quarks.  The preselection and
optimized cuts chosen for the four jet analysis are applied without
adjustment. The efficiency on
$\hZ\!\rightarrow\!\A\A\Z\!\rightarrow\!\bbbar\bbbar\qqbar$ is above
40\% over the region of interest.  
The mass $\chi^2$ of the four jet analysis is less effective in the six jet
topology,
however the $\Btag$ gives the final variable enough discriminating power
to distinguish between signal and background.
%The mass dependent final discriminant
%of the four jet 
%analysis, is no longer well defined
%signal from background and the analysis relies heavily on the b-tag
%variable.

%
%%%%%%%%%%%%%%%%%%%%%%%%%%%%%%%%%%%%%%%%%%%%%%%%%%%%%%%%%%%%%%%%%%%%%%%%%%%%%%%
% bb~\tau\tau Analysis
%%%%%%%%%%%%%%%%%%%%%%%%%%%%%%%%%%%%%%%%%%%%%%%%%%%%%%%%%%%%%%%%%%%%%%%%%%%%%%%
%

%\subsection*{$\boldsymbol{\hAtobbtt(\tautau\bbbar)}$,\, $\boldsymbol{\hZtobbtt}$\, and $\boldsymbol{\hZtottqq}$ Channels}
\subsection*{The $\boldsymbol{\hA}\!\boldsymbol{\rightarrow}\!\boldsymbol{\bbbar}\pmb{\tau}^{\boldsymbol{+}}\pmb{\tau}^{\boldsymbol{\, -}}$, $\boldsymbol{\hZ}\!\boldsymbol{\rightarrow}\!\boldsymbol{\bbbar}\pmb{\tau}^{\boldsymbol{+}}\pmb{\tau}^{\boldsymbol{\, -}}$\, and $\boldsymbol{\hZ}\!\boldsymbol{\rightarrow}\!\pmb{\tau}^{\boldsymbol{+}}\pmb{\tau}^{\boldsymbol{\, -}}\boldsymbol{\qqbar}$ Channels}
\label{sec:bbtt}
The signatures of $\hAtobbtt$, \hZtobbtt or \hZtottqq
events\footnote{The $\hAtottbb$ is also considered.} are a pair of taus
accompanied by two hadronic jets.  The main
background comes from W-pair decays containing taus.  Two analyses are
optimized for the hZ and for the hA channels. The hZ analysis follows that
of the Standard Model Higgs search and is
described in detail in Reference~\cite{l3_sm_higgs_99_paper}. The
$\hAtobbtt$ selection is described in this letter. As
in the Standard Model Higgs search, two selections are performed, one based
on the tau identification (particle-based selection) and the other
relying more on the event kinematics (jet-based selection).

First a common preselection
%~\cite{l3_sm_higgs_99_paper} 
is applied to
both analyses, then cuts specific to each analysis are chosen.
The major difference in the \hA selection from that
% between the \hZ selection
%used in the Standard Model Higgs search
%and the selection in the \hA search 
of the \hZ analysis is the need for greater sensitivity to
lower Higgs masses.  To accomplish this, the cuts on opening angles of
the jet and tau pairs have been removed, and the invariant mass cuts on
the tau-tau
%, $\mathrm{m_{\tau\tau}}$, 
and jet-jet systems
%, $\mathrm{m_{qq}}$, 
have been relaxed.
% to lie between 5\GeV and 125\GeV.
To reject the increased background accepted by loosening
the selection, additional cuts are applied which exploit the
kinematics of the \hA events.  A cut is placed on the ratio of the sum
of the energies of the tau decay products to the sum of the jet energies.
%, $(\mathrm{\frac{E_{\tau1}^{dec}+E_{\tau2}^{dec}}{E_{jet1}+E_{jet2}}})$,
%to be less than one.
The magnitude of the missing momentum vector in the rest frame of the Higgs
%, $\mathrm{p_{miss}^*}$, 
%is required to be less than 40\GeV,
is restricted,
where the taus are expected to be back-to-back
resulting in a partial cancellation of the missing momentum vectors.
Finally, there is a requirement on the cosine of the production angle
of the Higgs boson with respect to the beam axis
%, $|cos\Theta|$, to be less than 0.8,
similar to that in the four-jet \hA analysis.  The selection
cuts chosen for both the particle- and jet-based selections are shown
in Table~\ref{tab:bbtt_cuts}.  The number of events observed, the
number expected from background processes, and the signal efficiency
for the \hA and \hZ analyses, after combining the particle- and jet-based
selections, are shown in Table~\ref{tab:bbtt_eff}.

\begin{figure}[htb]
  \begin{center}
    \includegraphics*[width=0.9\textwidth,bb=11 199 622 1016]{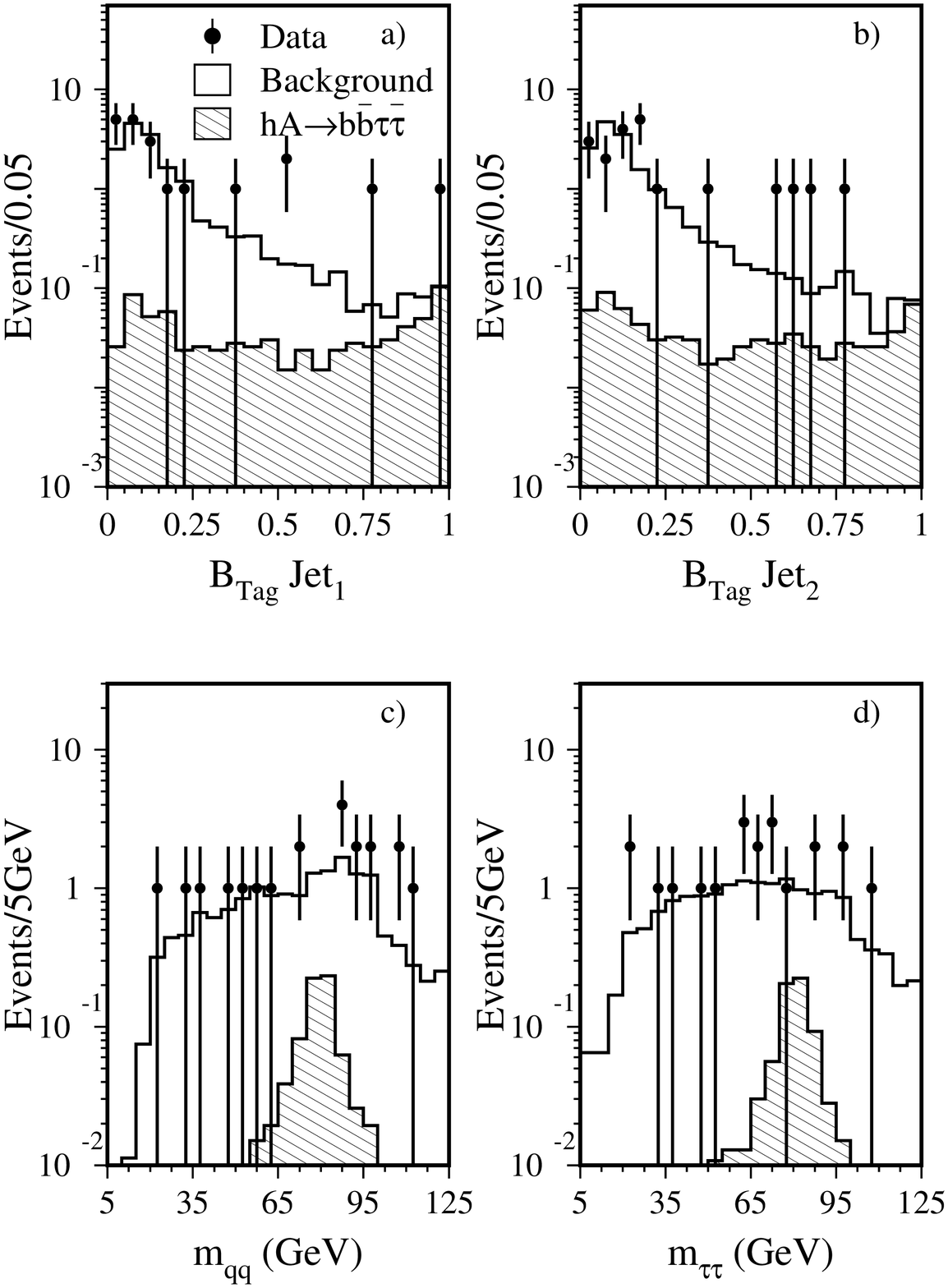}    
    \caption{The distributions for the $\hAtobbtt$
      search channel of a) the $\Btag$ for hadronic jet 1 and
      b) hadronic jet 2, c) the reconstructed mass for the hadronic
      system, and d) the reconstructed mass for the tau-tau system.
%      The points are the $\rts=189\GeV$ data, the open histograms the
%      Monte Carlo background and the hatched histograms the expected
      The hatched histogram is the
      $\hAtobbtt$ signal normalized for $\mh=80\GeV$ and $\tanb=50$.}
    \label{fig:last_vars_bbtt}
  \end{center}
\end{figure}

\begin{figure}[htb]
  \begin{center}
    \includegraphics*[width=0.9\textwidth,bb=9 205 669 1019]{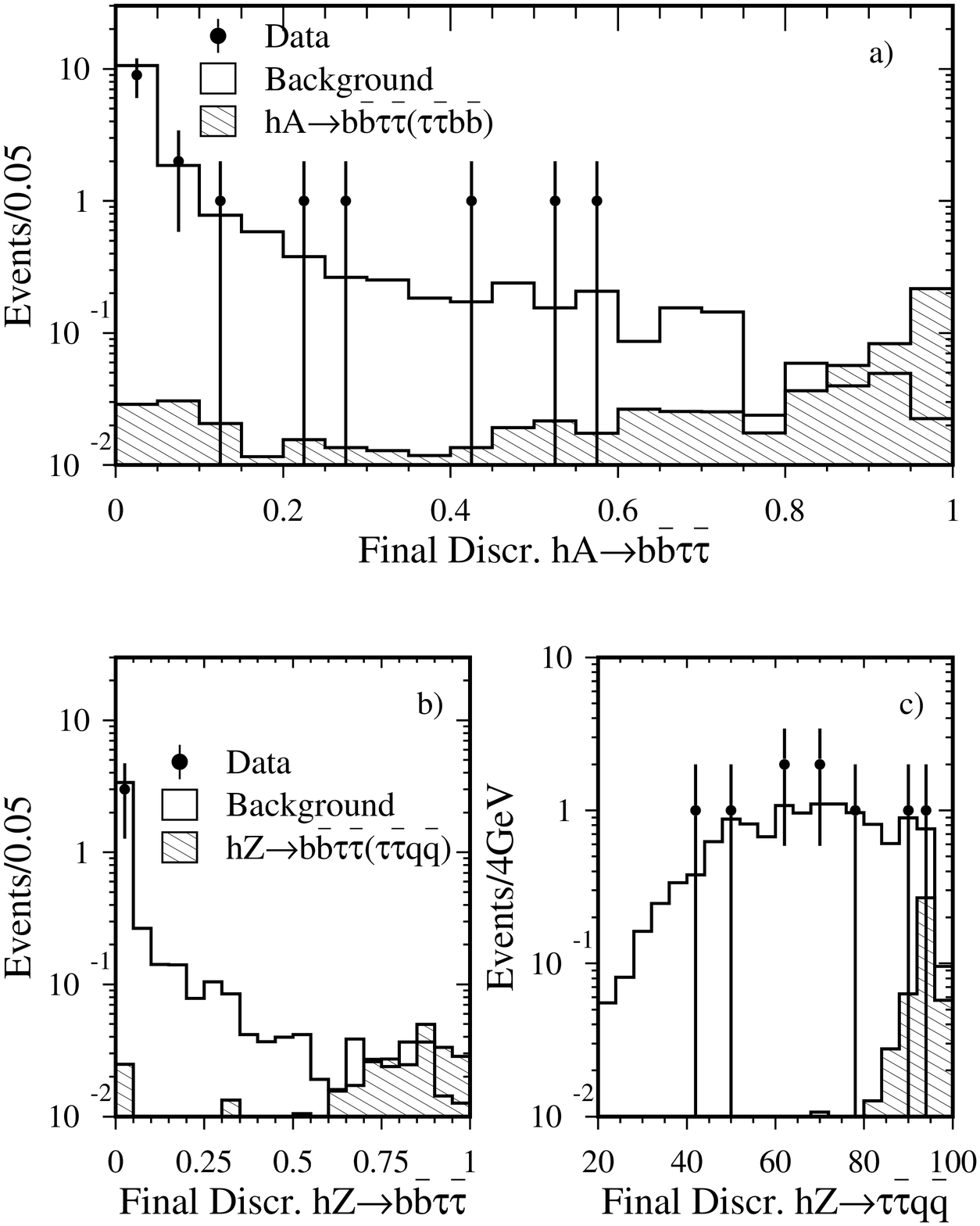}    
    \caption{Distributions of the final variables for the a)
      $\hAtobbtt$ for $\mh=80\GeV$ at $\tanb=50$,
      b) the
      \hZtobbtt for $\mh=95\GeV$ at $\tanb=3$ and c) the
      \hZtottqq search for $\mh=95\GeV$ at $\tanb=3$.}
    \label{fig:bbtt_final}
  \end{center}
\end{figure}

The final variable is the likelihood of the event to be \hA or \hZ
based on the $\Btag$ values for each hadronic jet, shown in
Figures~\ref{fig:last_vars_bbtt}a and~\ref{fig:last_vars_bbtt}b, and
the reconstructed invariant mass of either the jet or tau system,
shown in Figures~\ref{fig:last_vars_bbtt}c
and~\ref{fig:last_vars_bbtt}d, using the same technique as in the
Standard Model Higgs search.
%~\cite{l3_sm_higgs_99_paper}.
Events which pass the \hA as well as the \hZ selection are classified as
either \hA or \hZ depending on the cross section weighted values of
these likelihoods.  Examples of the final variable for the \hA search
at large values of \tanb and the \hZ search at low values of \tanb are
shown in Figure~\ref{fig:bbtt_final}.
\begin{table}[htbp]
  \begin{center}
    \leavevmode
    \begin{tabular}{cll} \hline
      Cut & \multicolumn{1}{c}{Particle-based selection} & \multicolumn{1}{c}{Jet-based selection} \\
      \hline
      Number of tracks & $\ge$ 5 & $\ge$ 5 \\
      Number of clusters & $\ge$ 15 & $\ge$ 15 \\
      $E_{\rm vis}/\sqrt{s}$ & $\ge$ 0.4, $\le$ 0.95 & $\ge$ 0.4, $\le$ 0.90\\
      $E_{\rm e}$,$E_{\mu}$,$E_{\gamma}$ & $\le$ 40\GeV & $\le$ 40\GeV \\
      ln\Ytf & $\ge$ -6 & $\ge$ -6 \\
      $E^{\tau}$
%      $(\mathrm{\frac{E_{\tau1}^{dec}+E_{\tau2}^{dec}}{E_{jet1}+E_{jet2}}})$
      & $\le$ 1 & $\le$ 1 \\
      $m_{\tau\tau}$,$m_{qq}$ & $\ge$ 5\GeV,$\le$ 125\GeV
      &$\ge$ 5\GeV,$\le$ 125\GeV \\
      $\mid\cos\Theta\mid$ & $\le$ 0.8 & $\le$ 0.8 \\
      $\mid p_{\rm miss}^* \mid$ & $\le$ 40\GeV & $\le$ 40\GeV \\
      $\mid\cos(\Theta_{\rm miss})\mid$ & - & $\le$ 0.95 \\
      $\alpha_{\mbox{\scriptsize \rm $\tau$-jet}}$ & - & $\ge$ 25$\,^\circ$ \\
      \hline
    \end{tabular}
    
    \caption{
      Selection cuts for particle-based and jet-based tau selections
      in the $\hAtobbtt$ search channel. In addition to those abbreviations
defined in the text: $E_{\rm vis}$ is the visible energy;
$E_{\rm e},\;E_{\mu}$ and $E_{\gamma}$ are the electron, muon and photon
energies, respectively; $E^{\tau}$ is the
ratio of the sum of the energies of the tau decay products to the sum of the
jet energies; $m_{\tau\tau}$,$m_{qq}$ is the invariant mass of the
tau-tau and jet-jet systems, respectively; $\Theta$ is the
production angle of
the Higgs boson with respect to the beam axis; $p_{\rm miss}^*$ is the
magnitude of the missing momentum vector in the rest frame of the Higgs;
$\Theta_{\rm miss}$ is the angle of missing energy vector with respect to
the beam axis; and $\alpha_{\mbox{\scriptsize \rm $\tau$-jet}}$ is the angle
between a tau jet and the closest quark jet.}
    \label{tab:bbtt_cuts}
  \end{center}
\end{table}

\begin{table}[htb]
  \begin{center}
%\hspace*{5mm}    
\begin{tabular}{crr}\hline
& \multicolumn{2}{c}{Number of Events} \\
Process & \hA selection & \hZ selection \\
\hline
\epemtoqq         & 2.3  & 2.3     \\
\epemtoWW         & 11.3 & 11.2    \\
\epemtoZZ         & 2.6  & 3.1     \\
$\mathrm{e^+e^-\rightarrow Ze^+e^-}$ & 0.4  & 0.5 \\
\hline
Total Expected      & 16.6  & 17.1   \\
Data        & 20 \hspace*{5.5pt}   & 12 \hspace*{5.5pt}   \\ \hline\hline
%    \end{tabular} \hspace*{5mm}
%    \begin{tabular}{lrr}
\hspace*{6pt}Efficiency \hAtobbtt \rule{0pt}{12pt}
            & 35.2\%   & 35.4\%  \\
%\hAtottbb   &          &         \\
\hspace*{1pt}Efficiency \hZtobbtt   & 21.1\%   & 30.0\%   \\
Efficiency \hZtottqq   & 21.8\% & 29.8\%  \\
\hline
    \end{tabular}
    \caption{Number of events expected and observed after selection
      for the tau search channels.  The signal
      efficiencies at $\rts=189\GeV$ are quoted for hA at $\mA=\mh=80\GeV$ and
      for hZ at $\mh=95\GeV$.}
    \label{tab:bbtt_eff}
  \end{center}
\end{table}

%
%%%%%%%%%%%%%%%%%%%%%%%%%%%%%%%%%%%%%%%%%%%%%%%%%%%%%%%%%%%%%%%%%%%%%%%%%%%%%%%
% Results
%%%%%%%%%%%%%%%%%%%%%%%%%%%%%%%%%%%%%%%%%%%%%%%%%%%%%%%%%%%%%%%%%%%%%%%%%%%%%%%
%
\subsection*{Results}
No evidence of the production of the \h and \A bosons is observed in
the data.  The excluded region of the MSSM parameter space
is evaluated by calculating the confidence level (\CL) that the
expected signal is absent in the observed data for the plane defined
by \tanbma.  The \CL is calculated using the technique described
in References~\cite{l3_1997_18,new_method}.  Bins of an analysis with a
signal-over-background ratio in the Monte Carlo of less than 0.05 are not considered in
the calculation of \CL.  This cut is chosen to minimize the effect of
systematic errors on the average \CL as calculated from a large set of
Monte Carlo trials.

Systematic errors on the signal and background are considered using
the same procedure as in the Standard Model Higgs
searches~\cite{l3_1997_18,l3_1998_11,l3_sm_higgs_99_paper}.
% taking into
% account detector, selection and theoretical uncertainties.
The overall
systematic error is estimated to be 5\% on the number of signal and
10\% on the number of background events. Statistical uncertainties due
to Monte Carlo statistics are completely
uncorrelated among the different bins of the individual channels and
have little effect on the final \CL calculation.  

\begin{figure}[htb]
  \begin{center}
    \includegraphics*[width=0.9\textwidth,bb=16 184 624 1038]{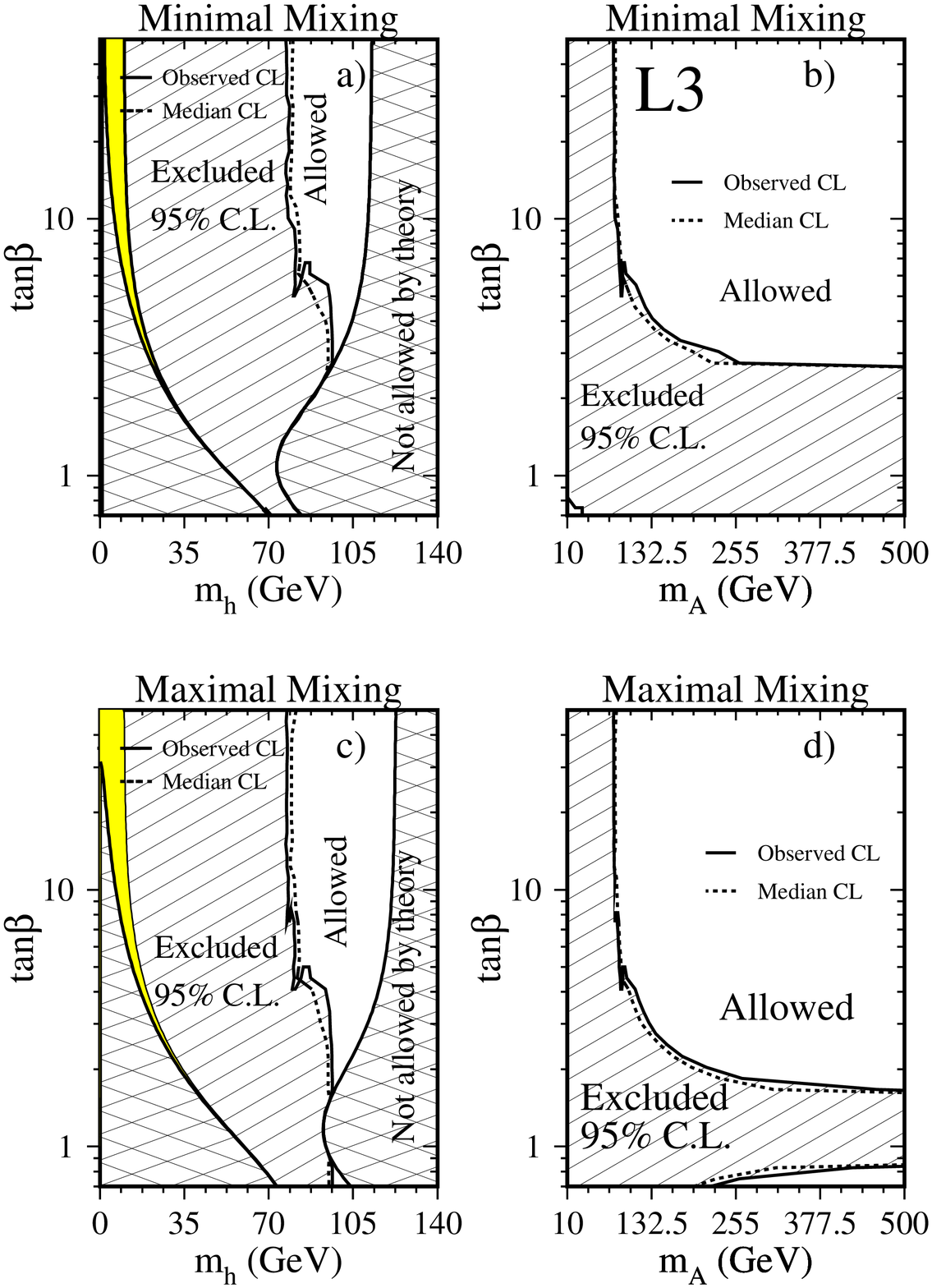}    
    \caption{Exclusion plots of the Higgs mass versus \tanb at the
      95\% \CL.  In all plots the area shaded by diagonal lines is the
      95\% exclusion, while the cross-hatched region is theoretically
      disallowed.  The grey region in plots a) and c) 
      corresponds to $\mA\!<\!10\GeV$ and has been previous excluded at
      LEP~\protect\cite{opal_2}.
      Plot a) is the
      95\% \CL exclusion of \mh versus \tanb in the minimal mixing
      scenario, and b) is the 95\% exclusion of \mA versus \tanb also for
      minimal mixing.  Plots c) and d) are the same for the maximal
      mixing scenario.}
    \label{fig:limit}
  \end{center}
\end{figure}

The data from the MSSM Higgs search using lower center-of-mass
energies~\cite{l3_1998_16} is combined with the $\rts=189\GeV$ data.
Figure~\ref{fig:limit} shows the region of the $(\tanb,\mh)$ plane and
the \tanbma plane excluded by L3 for the maximal and minimal mixing
scenarios.  On the plot, the 95\% \CL is shown as a solid line while
the expected median \CL is shown as a dashed line.
Table~\ref{tab:limit} lists the masses of the
\h and \A excluded at the 95\% \CL using the data at $\rts=189\GeV$
and lower center-of-mass energies for $\tanb=3$ and $\tanb=50$ as well
as the median and average exclusion and the probability to obtain a
higher limit.  The probability to obtain a higher limit reaches a
maximum in the high \tanb region with an \mh mass of 68\GeV, where
there is an upward fluctuation in the data. The lowest value of \mh
excluded is at $\tanb=15.0$ for maximal mixing and the lowest value of
\mA is excluded at $\tanb=50.0$ for minimal mixing.  An interesting
feature of these results is that the region of $0.8<\tanb<1.5$ is excluded
in the
MSSM, according to the current theoretical calculation of the maximum
Higgs mass allowed and for $m_{\rm top}$ equal to 175\GeV~\cite{hzha}.
However, recent two-loop
calculations~\cite{hollik98} seem to favor larger values of the
maximum allowed \mh in this region, which would change the excluded
band of \tanb.

For the MSSM parameters considered and assuming \tanb greater than
one, this results in lower mass limits at the 95\% \CL of

\begin{displaymath}
  \mh > 77.1 \GeV, \;\;
  \mA > 77.1 \GeV.
\end{displaymath}

\begin{table}[htbp]
  \begin{center}
    \leavevmode
    \begin{tabular}{c|cc|cccc|c} \hline
      & \multicolumn{6}{|c|}{Lower mass limits in\GeV at 95\% \CL} \\ \cline{2-7}
%      Mixing, \tanb & \mhavg & \mAavg & \mhmed & \mAmed & \CLb & $\boldsymbol{\mh}$ & $\boldsymbol{\mA}$  \\
%      \hline
%      minimal, 3  & 92.7 & 164.0 & 94.6 & 192.6 & 12\% & {\bf 96.3} & {\bf 225.0}  \\
%      minimal, 50 & 78.2 & 78.2 & 80.0 & 80.0 & 80\% & {\bf 77.1} & {\bf 77.1}  \\
%      maximal, 3  & 89.0 & 111.9 & 90.4 & 117.1 & 15\% & {\bf 95.4} & {\bf 128.9} \\
%      maximal, 50 & 78.9 & 79.0 & 81.4 & 81.5 & 77\% & {\bf 77.5} & {\bf 77.6} \\
      & \multicolumn{2}{|c|}{Observed} & \multicolumn{4}{|c|}{Expected} & \\
      Mixing, \tanb & $\boldsymbol{\mh}$ & $\boldsymbol{\mA}$ & \mhavg & \mAavg & \mhmed & \mAmed & \CLb  \\
      \hline
      minimal, 3  & {\bf 96.3} & {\bf 225.0} & 92.7 & 164.0 & 94.6 & 192.6 & 12\%  \\
      minimal, 50 & {\bf 77.1} & {\bf 77.1} & 78.2 & 78.2 & 80.0 & 80.0 & 80\%  \\
      maximal, 3  & {\bf 95.4} & {\bf 128.9} & 89.0 & 111.9 & 90.4 & 117.1 & 15\%  \\
      maximal, 50 & {\bf 77.5} & {\bf 77.6} & 78.9 & 79.0 & 81.4 & 81.5 & 77\% \\

      \hline
    \end{tabular}
    \caption{Higgs mass limits in the MSSM from the data at
      $\rts=130\GeV-189\GeV$.  The
      masses in boldface are the lower mass limits set at the
      \mbox{95\% \CL} from the data.  The
      masses $< \kern -0.4em m \kern -0.4em >$ and $\overline{m}$
      are respectively the average and median mass limits for the \h
      and \A bosons as calculated from a large set of Monte Carlo
      trials.  Assuming there is no signal, \CLb is the probability to
      obtain a mass limit on \mh larger than the one observed.
    \label{tab:limit}}
  \end{center}
\end{table}

%
%%%%%%%%%%%%%%%%%%%%%%%%%%%%%%%%%%%%%%%%%%%%%%%%%%%%%%%%%%%%%%%%%%%%%%%%%%%%%%%
% Acknowledgements
%%%%%%%%%%%%%%%%%%%%%%%%%%%%%%%%%%%%%%%%%%%%%%%%%%%%%%%%%%%%%%%%%%%%%%%%%%%%%%%
%
\section*{Acknowledgements}
We acknowledge the efforts of the engineers and technicians who
have participated in the construction and maintenance of L3 and
express our gratitude to the CERN accelerator divisions for the superb
performance of LEP.

%
%%%%%%%%%%%%%%%%%%%%%%%%%%%%%%%%%%%%%%%%%%%%%%%%%%%%%%%%%%%%%%%%%%%%%%%%%%%%%%%
% Bibliography
%%%%%%%%%%%%%%%%%%%%%%%%%%%%%%%%%%%%%%%%%%%%%%%%%%%%%%%%%%%%%%%%%%%%%%%%%%%%%%
%
% Style file to use with mcite.
% Use l3style with just cite.
%\bibliographystyle{/l3/paper/biblio/l3stylem}

%\vfill
%These L3 Internal Notes are freely available upon request from 
%The L3 Secretariat, CERN, CH--1211 Geneva 23, Switzerland. 
%Internet: http://l3www.cern.ch.
%
%%%%%%%%%%%%%%%%%%%%%%%%%%%%%%%%%%%%%%%%%%%%%%%%%%%%%%%%%%%%%%%%%%%%%%%%%%%%%%
% Author List
%%%%%%%%%%%%%%%%%%%%%%%%%%%%%%%%%%%%%%%%%%%%%%%%%%%%%%%%%%%%%%%%%%%%%%%%%%%%%%
%
\newpage
\typeout{   }     
\typeout{Using author list for paper 192 -?}
\typeout{$Modified: Mon Oct  4 14:47:07 1999 by clare $}
\typeout{!!!!  This should only be used with document option a4p!!!!}
\typeout{   }
%
%
%
%  L A T E X  version!!
%
%
% Make sure that the Lep package has been used!
%\input{Lep.sty}%
%
%\ifx\LepCalled\undefined%
%\typeout{     }%
%\typeout{!!!!!!!!!!!!!!!!!!!!!!!!!!!!!!!!!!!!!!!!!!!!!!!!!!!!!!!!!!!}%
%\typeout{Yikes.  You haven't used the Lep package!}%
%\typeout{Please put \protect\usepackage\protect{Lep\protect} in your preamble,
%         followed by}%
%\typeout{\protect\Lep\protect{1\protect} or \protect\Lep\protect{2\protect}}%
%\typeout{     }%
%\typeout{For now you will get a Lep phase 2 authorlist (may not be right!).}%
%\typeout{!!!!!!!!!!!!!!!!!!!!!!!!!!!!!!!!!!!!!!!!!!!!!!!!!!!!!!!!!!!}%
%\typeout{     }%
%\Lep{2}\fi%

\newcount\tutecount  \tutecount=0
\def\tutenum#1{\global\advance\tutecount by 1 \xdef#1{\the\tutecount}}
\def\tute#1{$^{#1}$}
\tutenum\aachen            % 1
\tutenum\nikhef            % 2
\tutenum\mich              % 3
\tutenum\lapp              % 4
\tutenum\basel             % 5
\tutenum\lsu               % 6
\tutenum\beijing           % 7
\tutenum\berlin            % 8
\tutenum\bologna           % 9 
\tutenum\tata              % 10
\tutenum\ne                % 11
\tutenum\bucharest         % 12
\tutenum\budapest          % 13
\tutenum\mit               % 14 
\tutenum\debrecen          % 15
\tutenum\florence          % 16
\tutenum\cern              % 17 
\tutenum\wl                % 18 
\tutenum\geneva            % 19
\tutenum\hefei             % 20
\tutenum\seft              % 21
\tutenum\lausanne          % 22
\tutenum\lecce             % 23
\tutenum\lyon              % 24
\tutenum\madrid            % 25
\tutenum\milan             % 26
\tutenum\moscow            % 27
\tutenum\naples            % 27
\tutenum\cyprus            % 29
\tutenum\nymegen           % 30
\tutenum\caltech           % 31
\tutenum\perugia           % 32
\tutenum\cmu               % 33
\tutenum\prince            % 34
\tutenum\rome              % 35
\tutenum\peters            % 36
\tutenum\salerno           % 37
\tutenum\ucsd              % 38
\tutenum\santiago          % 39
\tutenum\sofia             % 40
\tutenum\korea             % 41
\tutenum\alabama           % 42
\tutenum\utrecht           % 43
\tutenum\purdue            % 44
\tutenum\psinst            % 45
\tutenum\zeuthen           % 46
\tutenum\eth               % 47
\tutenum\hamburg           % 48
\tutenum\taiwan            % 49
\tutenum\tsinghua          % 50
{
\parskip=0pt
\noindent
{\bf The L3 Collaboration:}
\ifx\selectfont\undefined%  old style font selection
 \baselineskip=10.8pt
 \baselineskip\baselinestretch\baselineskip
 \normalbaselineskip\baselineskip
 \ixpt
\else%                      new style font selection
 \fontsize{9}{10.8pt}\selectfont
\fi
\medskip
\tolerance=10000
\hbadness=5000
\raggedright
\hsize=162truemm\hoffset=0mm
\def\r{\rlap,}
\noindent

M.Acciarri\r\tute\milan\
P.Achard\r\tute\geneva\ 
O.Adriani\r\tute{\florence}\ 
M.Aguilar-Benitez\r\tute\madrid\ 
J.Alcaraz\r\tute\madrid\ 
G.Alemanni\r\tute\lausanne\
J.Allaby\r\tute\cern\
A.Aloisio\r\tute\naples\ 
M.G.Alviggi\r\tute\naples\
G.Ambrosi\r\tute\geneva\
% ...
\newpage
%\rule{\textwidth}{0.4pt}
\begin{list}{A}{\itemsep=0pt plus 0pt minus 0pt\parsep=0pt plus 0pt minus 0pt
                \topsep=0pt plus 0pt minus 0pt}
\item[\aachen]
 I. Physikalisches Institut, RWTH, D-52056 Aachen, FRG$^{\S}$\\
 III. Physikalisches Institut, RWTH, D-52056 Aachen, FRG$^{\S}$
\item[\nikhef] National Institute for High Energy Physics, NIKHEF, 
     and University of Amsterdam, NL-1009 DB Amsterdam, The Netherlands
\item[\mich] University of Michigan, Ann Arbor, MI 48109, USA
\item[\lapp] Laboratoire d'Annecy-le-Vieux de Physique des Particules, 
     LAPP,IN2P3-CNRS, BP 110, F-74941 Annecy-le-Vieux CEDEX, France
\item[\basel] Institute of Physics, University of Basel, CH-4056 Basel,
     Switzerland
\item[\lsu] Louisiana State University, Baton Rouge, LA 70803, USA
\item[\beijing] Institute of High Energy Physics, IHEP, 
  100039 Beijing, China$^{\triangle}$ 
\item[\berlin] Humboldt University, D-10099 Berlin, FRG$^{\S}$
\item[\bologna] University of Bologna and INFN-Sezione di Bologna, 
     I-40126 Bologna, Italy
\item[\tata] Tata Institute of Fundamental Research, Bombay 400 005, India
\item[\ne] Northeastern University, Boston, MA 02115, USA
\item[\bucharest] Institute of Atomic Physics and University of Bucharest,
     R-76900 Bucharest, Romania
\item[\budapest] Central Research Institute for Physics of the 
     Hungarian Academy of Sciences, H-1525 Budapest 114, Hungary$^{\ddag}$
\item[\mit] Massachusetts Institute of Technology, Cambridge, MA 02139, USA
\item[\debrecen] KLTE-ATOMKI, H-4010 Debrecen, Hungary$^\P$
\item[\florence] INFN Sezione di Firenze and University of Florence, 
     I-50125 Florence, Italy
\item[\cern] European Laboratory for Particle Physics, CERN, 
     CH-1211 Geneva 23, Switzerland
\item[\wl] World Laboratory, FBLJA  Project, CH-1211 Geneva 23, Switzerland
\item[\geneva] University of Geneva, CH-1211 Geneva 4, Switzerland
\item[\hefei] Chinese University of Science and Technology, USTC,
      Hefei, Anhui 230 029, China$^{\triangle}$
\item[\seft] SEFT, Research Institute for High Energy Physics, P.O. Box 9,
      SF-00014 Helsinki, Finland
\item[\lausanne] University of Lausanne, CH-1015 Lausanne, Switzerland
\item[\lecce] INFN-Sezione di Lecce and Universit\'a Degli Studi di Lecce,
     I-73100 Lecce, Italy
\item[\lyon] Institut de Physique Nucl\'eaire de Lyon, 
     IN2P3-CNRS,Universit\'e Claude Bernard, 
     F-69622 Villeurbanne, France
\item[\madrid] Centro de Investigaciones Energ{\'e}ticas, 
     Medioambientales y Tecnolog{\'\i}cas, CIEMAT, E-28040 Madrid,
     Spain${\flat}$ 
\item[\milan] INFN-Sezione di Milano, I-20133 Milan, Italy
\item[\moscow] Institute of Theoretical and Experimental Physics, ITEP, 
     Moscow, Russia
\item[\naples] INFN-Sezione di Napoli and University of Naples, 
     I-80125 Naples, Italy
\item[\cyprus] Department of Natural Sciences, University of Cyprus,
     Nicosia, Cyprus
\item[\nymegen] University of Nijmegen and NIKHEF, 
     NL-6525 ED Nijmegen, The Netherlands
\item[\caltech] California Institute of Technology, Pasadena, CA 91125, USA
\item[\perugia] INFN-Sezione di Perugia and Universit\'a Degli 
     Studi di Perugia, I-06100 Perugia, Italy   
\item[\cmu] Carnegie Mellon University, Pittsburgh, PA 15213, USA
\item[\prince] Princeton University, Princeton, NJ 08544, USA
\item[\rome] INFN-Sezione di Roma and University of Rome, ``La Sapienza",
     I-00185 Rome, Italy
\item[\peters] Nuclear Physics Institute, St. Petersburg, Russia
\item[\salerno] University and INFN, Salerno, I-84100 Salerno, Italy
\item[\ucsd] University of California, San Diego, CA 92093, USA
\item[\santiago] Dept. de Fisica de Particulas Elementales, Univ. de Santiago,
     E-15706 Santiago de Compostela, Spain
\item[\sofia] Bulgarian Academy of Sciences, Central Lab.~of 
     Mechatronics and Instrumentation, BU-1113 Sofia, Bulgaria
\item[\korea] Center for High Energy Physics, Adv.~Inst.~of Sciences
     and Technology, 305-701 Taejon,~Republic~of~{Korea}
\item[\alabama] University of Alabama, Tuscaloosa, AL 35486, USA
\item[\utrecht] Utrecht University and NIKHEF, NL-3584 CB Utrecht, 
     The Netherlands
\item[\purdue] Purdue University, West Lafayette, IN 47907, USA
\item[\psinst] Paul Scherrer Institut, PSI, CH-5232 Villigen, Switzerland
\item[\zeuthen] DESY, D-15738 Zeuthen, 
     FRG
\item[\eth] Eidgen\"ossische Technische Hochschule, ETH Z\"urich,
     CH-8093 Z\"urich, Switzerland
\item[\hamburg] University of Hamburg, D-22761 Hamburg, FRG
\item[\taiwan] National Central University, Chung-Li, Taiwan, China
\item[\tsinghua] Department of Physics, National Tsing Hua University,
      Taiwan, China
\item[\S]  Supported by the German Bundesministerium 
        f\"ur Bildung, Wissenschaft, Forschung und Technologie
\item[\ddag] Supported by the Hungarian OTKA fund under contract
numbers T019181, F023259 and T024011.
\item[\P] Also supported by the Hungarian OTKA fund under contract
  numbers T22238 and T026178.
\item[$\flat$] Supported also by the Comisi\'on Interministerial de Ciencia y 
        Tecnolog{\'\i}a.
\item[$\sharp$] Also supported by CONICET and Universidad Nacional de La Plata,
        CC 67, 1900 La Plata, Argentina.
\item[$\diamondsuit$] Also supported by Panjab University, Chandigarh-160014, 
        India.
\item[$\triangle$] Supported by the National Natural Science
  Foundation of China.
\item[\dag] Deceased.
\end{list}
}
\vfill

%%% Local Variables: 
%%% mode: latex
%%% TeX-master: t
%%% End:

%%% Local Variables: 
%%% mode: latex
%%% TeX-master: t
%%% TeX-master: t
%%% TeX-master: t
%%% End: 

%%% Local Variables: 
%%% mode: latex
%%% TeX-master: t
%%% End: 

%%% Local Variables: 
%%% mode: latex
%%% TeX-master: t
%%% End: 

%%% Local Variables: 
%%% mode: latex
%%% TeX-master: t
%%% End: 

\newpage

\end{document}